\documentclass[%
twocolumn
, hidempi
]{mpi2015-cscpreprint}

\usepackage{amsmath}
\usepackage{amssymb}

\usepackage{algorithmic}
\usepackage{algorithm}

\usepackage{amsfonts,mathrsfs}

\usepackage{mathptmx}       % selects Times Roman as basic font
\usepackage{helvet}         % selects Helvetica as sans-serif font
\usepackage{courier}        % selects Courier as typewriter font
\usepackage{type1cm}        % activate if the above 3 fonts are
\usepackage{makeidx}         % allows index generation
\usepackage{graphicx}        % standard LaTeX graphics tool
\usepackage{color}

\usepackage{amsmath,amssymb,amsbsy,latexsym,bbold}
\usepackage{mathrsfs}
\usepackage{graphicx}
\usepackage{epstopdf}
\usepackage{multirow}
\usepackage{enumerate}
\usepackage{cite}

\makeindex             % used for the subject index
\newcommand{\la}{ \lambda }
\newcommand{\ql}{ q^{(l)} }
\newcommand{\qr}{ q^{(r)} }

\newcommand{\bv}{{v}}
\newcommand{\bw}{{w}}

\newcommand{\cO}{ {\cal O} }

\newcommand{\bA}{{ A}}
\newcommand{\bB}{{ B}}
\newcommand{\bC}{{ C}}

\newcommand{\bE}{{ E}}

\newcommand{\bX}{{ X}}
\newcommand{\bY}{{ Y}}

\newcommand{\bfe}{{\mathbf e}}
\newcommand{\br}{{\mathbf r}}
\newcommand{\bu}{{\mathbf u}}
\newcommand{\bx}{{\mathbf x}}

\newcommand{\by}{{\mathbf y}}
\newcommand{\bl}{{ l}}
\renewcommand{\br}{{ r}}

\def\IR{{\mathbb R}}
\def\IC{{\mathbb C}}
\newcommand{\cR}{ {\cal R} }
\def\IL{{\mathbb L}}
\def\IV{{\mathbb V}}
\def\IW{{\mathbb W}}
\newcommand{\sIL}{{{{\mathbb L}_s}}}

\newcommand{\Si}{{\boldsymbol{\Sigma}}}

\providecommand{\Htran}[0]{H} %
\providecommand{\rank}[0]{\mathrm{rank}} %
 \providecommand{\ie}[0]{\textit{i.e.}~} %

\newtheorem{Definition}{Definition}

\newcommand{\QNUM}{n_{\mathrm p}}

\newtheorem{theorem}{Theorem}
\newtheorem{remark}{Remark}

\newtheorem{example}{Example}
\newtheorem{Assumption}{Assumption}

\newcommand{\NX}{n_\mathrm{x}}
\newcommand{\NY}{n_\mathrm{y}}
\newcommand{\NU}{n_\mathrm{u}}

\newcommand{\X}{\mathbb{R}^{n_\mathrm{x}}}

\newcommand{\AQ}{\mathbb{I}_0^{n_\mathrm{p}}}

\begin{document}
	
\title{Reduced-order modeling of LPV systems in the Loewner framework}

\author[$\ast$]{Ion Victor Gosea}
\affil[$\ast$]{Max Planck Institute for Dynamics of Complex Technical Systems,
	Sandtorstr. 1, 39106 Magdeburg, Germany.\authorcr
	\email{gosea@mpi-magdeburg.mpg.de}, \orcid{0000-0003-3580-4116}}

\author[$\dagger$]{Mihaly Petreczky}
\affil[$\dagger$]{Centre de Recherche en Informatique, Signal et Automatique de Lille (CRIStAL), UMR CNRS 9189, CNRS Lille, France.\authorcr
	\email{mihaly.petreczky@ec-lille.fr}, \orcid{0000-0003-2264-5689}}

\author[$\ddagger$]{Athanasios C. Antoulas}
\affil[$\ddagger$]{Rice University, Houston, Texas, USA, and Max Planck Institute for Dynamics of Complex Technical Systems, Magdeburg, Germany.\authorcr
	\email{aca@rice.edu}, }

\shorttitle{Data-driven MOR of LPV systems}
\shortauthor{I. V. Gosea, M. Petreczky, A.C. Antoulas}
\shortdate{}

%\keywords{\small model reduction, parametric systems, rational interpolation, Sylvester equations, holomorphic functions. \normalsize}

%\msc{MSC1, MSC2, MSC3}

\abstract{%
We propose a model reduction method for LPV systems. We consider LPV state-space representations with an affine dependence on the scheduling variables. The main idea behind the proposed method is to compute the reduced order model in such a manner that its frequency domain transfer function coincides with that of the original model for some frequencies. The proposed method uses Loewner-like matrices, which can be calculated from the frequency domain representation of the system. The contribution of the paper represents an extension of the well-established Loewner framework to LPV models.}

%     \novelty{This work proposes a new MOR method that .}  

\maketitle

\section{Introduction}

\emph{Linear parameter-varying} (LPV) systems are linear  systems where the coefficients are  functions of a time-varying signal, the so-called \emph{scheduling variable}.  %Such equations are called LPV systems \cite{Toth2010SpringerBook,Toth11_LPVBehav}. 
%That is, LPV systems are a class of mathematical models having a certain structure (linear and time-varying).
Control design and system identification of LPV systems is a popular topic \cite{Rugh00,LPVBook2012,Val13b,Toth2010SpringerBook,toth12,Giarre2002,Wingerden09,LopesDosSantos2008,Sznaier:01,Verdult02,BlanchiniTAC}.
Model reduction refers to a general class of methodologies used to reduce the complexity of typically large-scale models, by approximating them with simpler, smaller models (and by retaining, at the same time, the main characteristics of the original model). We refer the reader to \cite{Ant05,BGW15surveyMOR,AntBG20}, and to the references therein for more details on some of the recent methods developed. Model reduction has also been investigated for LPV systems in the last two decades; we refer the reader to the collection  \cite{farhood2003, dehillerin2011, adegas2013, wood1996, widowatin,MertBastug:CDC2015,toth12,BennerLPV,TothModRed,SzaboModelRed,WeilandModelRed}, for more details. However, model reduction of LPV systems preserving some component of the frequency response has not been investigated so far, to the best of our knowledge.

In this paper we propose a model reduction method which preserves some component of the frequency response of an LPV model. 
We will concentrate on LPV state-space representations with an affine dependence on the scheduling parameters. This approach is an extension of the well-known Loewner framework for LTI systems \cite{MA07}  and it is closely related to the Loewner framework for linear switched systems \cite{VictorLoewner} and bilinear systems \cite{AGI16}. 
The basic idea is to define a set of \emph{generalized transfer functions}  which 
represent the multivariate Laplace transforms of the input-output map of an LPV system. 
The definition of these generalized transfer functions resembles that of bilinear systems \cite{Rugh96book}, and it is closely related to generalized kernel functions for linear switched systems \cite{VictorLoewner}. Similarly, the ensuing Loewner framework formulated here for LPV systems follows closely that for 
linear switched systems \cite{VictorLoewner}, and bears some resemblance with that for bilinear systems \cite{AGI16}.

The motivation for formulating a moment matching model reduction algorithm for LPV systems is as follows. First, it allows to deal with LPV systems which are not quadratically stable. This is in contrast to model reduction methods based on balanced truncation
or solving LMIs \cite{farhood2003, dehillerin2011, adegas2013, wood1996,BennerLPV,WeilandModelRed}, and its computation complexity is likely to be lower than that of methods based on solving LMIs.  Second, it has a system theoretic interpretation
in frequency domain.
%in contrast to methods based on discarding 
%certain singular  values of the Hankel-matrix \cite{toth12}.
Finally, in contrast to moment matching methods based on matching sub-Markov parameters \cite{MertBastug:CDC2015}, the input-output behavior of the reduced model is an approximation of the original one for scheduling signals and control inputs which are  linear combinations of certain harmonics. That is, it is possible to relate the frequency response of the original and reduced model. In turn, for LPV control synthesis the use of frequency domain specifications is quite natural, rendering the model reduction method compatible with control design.

To the best of our knowledge, the results of the paper are new. 
The existing literature is mostly applicable for stable LPV systems. 
%except \cite{widowati,toth12,MertBastug:CDC2015} they are only applicable to quadratically stable LPV systems. 
The method of \cite{widowatin} is applicable to quadratically stabilizable and
detectable LPV systems.
In contrast, this paper does not impose any stability restrictions on the class of LPV systems. 
In \cite{toth12}, a modification of the realization algorithm is proposed. However, it requires the construction of the Hankel matrix and hence it suffers from the curse of dimensionality. 
In \cite{SirajCDC2012}, reduction of the number of states and the number of scheduling parameters  was investigated. However, the method of \cite{SirajCDC2012} requires constructing the Hankel matrix explicitly. Hence, it displays the same type of challenges as the method in \cite{toth12}.
%In addition, the system theoretic interpretation of the algorithm is less clear. 

\textbf{Outline:}
In Section \ref{sec:pre} we present the definition of the model class, their input-output maps, equivalence and minimality, following \cite{petreczky2016}.
In Section \ref{sect:gentransfer}, the definition of generalized transfer functions for LPV models is presented. In Section \ref{sect:lti} contains a  brief introduction to the classical Loewner framework  for LTI systems. Section \ref{sect:main}
contains the presentation of the main result. In Section \ref{sect:num} we present a numerical example to illustrate the proposed model reduction method.

\section{Preliminaries} \label{sec:pre}
\subsection{Notation and terminology}
Let $\mathbb{N}$ be the set of all natural numbers including zero. 
For a finite set $X$, denote by $\mathcal{S}(X)$ the set of finite sequences generated by elements from $X$, \emph{i.e.}, each $s \in \mathcal{S}(X)$ is of the form $s=\zeta_{1}\zeta_{2} \cdots \zeta_{k}$ with $\zeta_1,\zeta_2,\ldots,\zeta_k \in X$, $k\in\mathbb{N}$;
$|s|$ denotes the length of the sequence $s$. For $s,r \in \mathcal{S}(X)$, $sr\in \mathcal{S}(X)$ denotes the concatenation of $s$ and $r$.  The symbol $\varepsilon$ is used for the empty sequence and $|\varepsilon|=0$ with $s\varepsilon=\varepsilon s=s$.
Denote by  $X^{\mathbb{N}}$ the set of all functions of the form $f:\mathbb{N} \rightarrow X$. 
%For each $j=1,\ldots,m$, $e_j$ is the $j^\mathrm{th}$ \emph{standard basis} in $\mathbb{R}^{m}$.
Let $\mathbb{I}_{\tau_1}^{\tau_2}=\{s\in\mathbb{Z}\mid  \tau_1 \leq s\leq \tau_2\}$ be an index set.
%Let $\delta_{i,j}$ be the Kronecker symbol,
%Before going further, some notations, described usually in automata theory \cite{Eil74}, must be introduced. More specifically, consider a (possibly infinite) set $X$. Denote by $X^{*}$ the set of finite sequences of elements of $X$, \emph{i.e.}, each $w \in X^{*}$ is of the form $w=a_{1}a_{2} \cdots a_{k}$, $a_1,a_2,\ldots,a_k \in X$, $t > 0$. The length of the sequence $w$  above is denoted by $|w|$. We denote by $wv$ the concatenation of the sequences $w,v \in X^{*}$, 

Let $\mathbb{T}=\mathbb{R}_{0}^{+}=[0,+\infty)$ be the continuous-time time axis. 
%Note that in both cases we exclude negative time instances. 
%Denote by \emph{$\xi$ the differentiation operator $\frac{d}{dt}$ (in CT) and the forward time-shift
%	operator $q$ (in DT)}, \emph{i.e.},
%if $z: \mathbb{T} \rightarrow \mathbb{R}^{n}$,  then
%$(\xi z)(t)=\frac{d}{dt} z(t)$, if $\mathbb{T}=\mathbb{R}_{0}^{+}$, and
%$(\xi z)(t)=z(t+1)$, if $\mathbb{T}=\mathbb{N}$.
%Denote by $\xi^k$ the $k$-fold application of $\xi$, \emph{i.e.}, for any $z:\mathbb{T} \rightarrow \mathbb{R}^n$,
%$\xi^{0} z=z$, and $\xi^{k+1} z = \xi (\xi^k z)$ for all $k \in \mathbb{N}$.
%Both for CT and DT, for any $\tau \in \mathbb{T}$, define the
%time shift operator $q^{\tau}$ as follows: for any $f:\mathbb{T} \rightarrow \mathbb{R}^n$, $q^{\tau}f:\mathbb{T} \rightarrow \mathbb{R}^n$ is
%defined by $(q^{\tau}f)(t)=f(t+\tau)$, $t \in \mathbb{T}$.

A function $f:\mathbb{R}_{0}^{+} \rightarrow \mathbb{R}^{n}$ is called
\emph{piecewise-continuous}, if $f$ has finitely many points of
discontinuity on any compact subinterval of $\mathbb{R}_{0}^{+}$ and, at any point of
discontinuity, the left-hand and right-hand side limits of $f$ exist and 
are finite.  We denote by $\mathcal{C}_\mathrm{p}(\mathbb{R}_{0}^{+},\mathbb{R}^n)$ the set of all
\emph{piecewise-continuous functions} of the above form. We denote by
$\mathcal{C}_\mathrm{d}(\mathbb{R}_{0}^{+},\mathbb{R}^n)$ the set of all 
differentiable functions of the form $f:\mathbb{R}_{0}^{+} \rightarrow \mathbb{R}^{n}$. 

\subsection{System theoretic definitions}
%In this paper, we consider the class of LPV systems that have 
An LPV \emph{state-space} (SS) representation with \emph{affine} linear dependence on the \emph{scheduling variable} (abbreviated as LPV-SSA) 
is a state-space representation of the form
%\footnote{Note that in our definition the output $y(t)$ at time $t$ does not depend on the input at time $t$. This restriction is made in order to simplify notation and most of the results can be easily extended when a direct dependence of $y(t)$ on $u(t)$ is included.}
\begin{equation}\label{equ:alpvss}
\Sigma \ \left\{
\begin{array}{lcl}
\dot x (t) &=& A(p(t)) x(t) + B(p(t)) u(t), \\
% \sum_{q=1}^{D} (A_{q} x(t) + B_{q}u(t)) p_q(t)  \\
y(t)  &=& C(p(t)) x(t) + D(p(t))u(t),
% \sum_{q=1}^{D} (C_{q}x(t)) p_q(t)
\end{array}\right.
\end{equation}
where $x(t) \in \mathbb{X}=\mathbb{R}^{n_\mathrm{x}}$ is the state, $y(t) \in \mathbb{Y}=\mathbb{R}^{n_\mathrm{y}}$ is the output, $u (t) \in \mathbb{U}=\mathbb{R}^{n_\mathrm{u}}$is the input, and $p(t) \in \mathbb{P}\subseteq \mathbb{R}^{n_\mathrm{p}}$ is the value of the \emph{scheduling variable} at time $t$, and
$A,B, C,D$ are matrix valued functions on $\mathbb{P}$ defined as
\begin{equation}
\label{equ:affdep}
\begin{split} %\mathcal{P}=\mathbb{P}^\mathbb{N}
A(\mathbb{p}) = A_0 + \sum_{i=1}^{\QNUM} A_i \mathbb{p}_i \mbox{, \ \  } 
B(\mathbb{p}) = B_0 + \sum_{i=1}^{\QNUM} B_i \mathbb{p}_i,   \\
C(\mathbb{p}) = C_0 + \sum_{i=1}^{\QNUM} C_i \mathbb{p}_i \mbox{, \ \ } 
D(\mathbb{p}) = D_0 + \sum_{i=1}^{\QNUM} D_i \mathbb{p}_i,
\end{split}
\end{equation}
for every $\mathbb{p}=[\begin{array}{cccc} \mathbb{p}_1 & \mathbb{p}_2 & \cdots & \mathbb{p}_{\QNUM}\end{array}]^\top \in \mathbb{P}$, with constant matrices $A_i \in \mathbb{R}^{\NX \times \NX}$,
$B_i \in \mathbb{R}^{\NX \times \NU}$, $C_i \in \mathbb{R}^{\NY \times \NX}$ and
$D_i \in \mathbb{R}^{\NY \times \NU}$ for all $i\in\mathbb{I}_0^{n_\mathrm{p}}$. 
\emph{It is assumed that $\mathbb{P}$ contains an affine basis of $\mathbb{R}^{n_\mathrm{p}}$} (see \cite{WebsterBook} for the definition of an affine basis). 
%Intuitively, $p$ corresponds to varying-operating conditions, nonlinear/time-varying dynamical aspects and /or external effects influencing the plant behavior and it is allowed to vary in the  set $\mathbb{P}$, see \cite{Toth2010SpringerBook}. 
In the sequel, we use the tuple
\begin{equation*}
\Sigma=(\mathbb{P},\left\{ A_i, B_i, C_i, D_i \right\}_{i=0}^{\QNUM})
\end{equation*}
to denote an LPV-SSA of the form \eqref{equ:alpvss} and use  $\dim{( \Sigma )}=n_\mathrm{x}$ to denote its state dimension. 
Define 
$\mathcal{X}=\mathcal{C}_\mathrm{d}(\mathbb{R}_{0}^{+},\mathbb{X})$, $\mathcal{Y}=\mathcal{C}_\mathrm{p}(\mathbb{R}_{0}^{+},\mathbb{Y})$, $\mathcal{U}=\mathcal{C}_\mathrm{p}(\mathbb{R}_{0}^{+},\mathbb{U})$, $\mathcal{P}=\mathcal{C}_\mathrm{p}(\mathbb{R}_{0}^{+},\mathbb{P})$.
%and $\mathcal{X}=\mathbb{X}^\mathbb{N}$, $\mathcal{Y}=\mathbb{Y}^{\mathbb{N}}$, $\mathcal{U}=\mathbb{U}^{\mathbb{N}}$, $\mathcal{P}=\mathbb{P}^{\mathbb{N}}$ in DT.
By a solution  of $\Sigma$ we mean a tuple of trajectories $(x,y,u,p)\in(\mathcal{X},\mathcal{Y},\mathcal{U},\mathcal{P})$ such that
\eqref{equ:alpvss} holds for all $t \in \mathbb{T}$.
For an initial state $x_\mathrm{o} \in \mathbb{X}$ define the  
\emph{input-to-state map}  $\mathfrak{X}_{\Sigma,x_\mathrm{o}}$ and the
\emph{input-output}  map $\mathfrak{Y}_{\Sigma,x_{\mathrm o}}$  of $\Sigma$ induced by $x_{\mathrm o}$ as 
\begin{equation}
\mathfrak{X}_{\Sigma,x_\mathrm{o}}:   \mathcal{U} \times \mathcal{P}  \rightarrow \mathcal{X}, \quad
\mathfrak{Y}_{\Sigma,x_\mathrm{o}}:   \mathcal{U} \times \mathcal{P}  \rightarrow \mathcal{Y}, 
\end{equation}
such that for any $(x,y,u,p) \in \mathcal{X} \times \mathcal{Y} \times \mathcal{U} \times \mathcal{P}$,
$x=\mathfrak{X}_{\Sigma,x_\mathrm{o}}(u,p)$ and 
$y=\mathfrak{Y}_{\Sigma,x_\mathrm{o}}(u,p)$ holds if and only if
$(x,y,u,p)$ is a solution of \eqref{equ:alpvss} and $x(0)=x_{\mathrm o}$.

We say that $\Sigma$ is \emph{span-reachable} from an initial state $x_\mathrm{o} \in \mathbb{X}$, if  $\mathrm{Span}\{ \mathfrak{X}_{\Sigma,x_\mathrm{o}}(u,p)(t) \mid (u,p) \in \mathcal{U} \times \mathcal{P}, t \in \mathbb{T} \}\!\! =\!\! \mathbb{X}$. In this paper we will concentrate on zero initial states, hence we will 
say that $\Sigma$ is \emph{span-reachable}, if 
it is span-reachable from the zero initial state. We say that $\Sigma$ is \emph{observable}, if for any two initial states $\bar{x}_{\mathrm o},\hat{x}_{\mathrm o} \in \X$, $\mathfrak{Y}_{\Sigma,\hat{x}_{\mathrm o}} = \mathfrak{Y}_{\Sigma,\bar{x}_{\mathrm o}}$ implies $\hat{x}_{\mathrm o} = \bar{x}_{\mathrm o}$.
% Recall that $\mathbb{T}=\mathbb{N}$ in DT and $\mathbb{T}=\mathbb{R}_0$ in CT. 
%\end{definition}
%That is, observability means that for any two distinct states of the system, the resulting outputs will be  different for some input and scheduling signals. 
Let $\Sigma$ of the form \eqref{equ:alpvss}  and $\Sigma^\prime=(\mathbb{P}, \{ A_i^{'}, B_i^{'}, C_i^{'}, D_i^{'} \}_{i=0}^{\QNUM})$ with $\dim(\Sigma)=\dim(\Sigma^\prime)=n_\mathrm{x}$. A nonsingular matrix $T \in \mathbb{R}^{n_\mathrm{x} \times n_\mathrm{x}}$ is said to be an \emph{isomorphism} from $\Sigma$ to $\Sigma^\prime$, if %for all $i \in \mathbb{I}_0^{n_{\mathrm p}}$, 
\begin{equation*}
%\label{equ:isomorphism}
\forall i \in \AQ:  A^\prime_{i} T=T A_i,~  B^\prime_{i}=T B_i, ~ C^\prime_{i}T= C_i,~ D^\prime_i=D_i.
\end{equation*}
We formalize  the input-output behavior of LPV-SSAs as maps of the form
\begin{equation}\label{equ:iofunction}
\mathfrak{F} : \mathcal{U} \times \mathcal{P} \rightarrow \mathcal{Y}.
\end{equation}
While any input-output map of an LPV-SSA induced by some initial state is of the above form, the converse
is not true.
%However, not all maps of the form \eqref{equ:iofunction} arise as
%input-output maps of some LPV-SSA.
% Let $\mathfrak{F}$ be an input-output function. 
The LPV-SSA $\Sigma$ is a \emph{realization} of an input-output map $\mathfrak{F}$ of the form \eqref{equ:iofunction} from \emph{the initial state $x_{\mathrm o} \in \mathbb{X}$}, if  $\mathfrak{F}=\mathfrak{Y}_{\Sigma,x_\mathrm{o}}$. 
In this paper we will concentrate on LPV-SSA realizations from the zero initial state. Accordingly, we will say $\Sigma$ is \emph{realization} of $\mathfrak{F}$, if $\Sigma$ is a realization of $\mathfrak{F}$ from zero initial state. 
An LPV-SSA $\Sigma$ is a \emph{minimal realization of $\mathfrak{F}$ from the initial state $x_\mathrm{o}$}, if
$\Sigma$ is a realization of $\mathfrak{F}$ from the initial state $x_{\mathrm{o}}$, and 
%\begin{itemize}
%\item $\exists x_\mathrm{o}\in\mathbb{X}$ such that $\mathfrak{Y}_{\Sigma,x_\mathrm{o}}= \mathfrak{F}$.
for every LPV-SSA $\Sigma^{\prime}$  which is a realization of $\mathfrak{F}$, $\dim{ ( \Sigma ) } \leq \dim{ ( \Sigma^\prime ) }$.
Again, when the initial state is zero, we say that $\Sigma$ is a \emph{minimal realization} of $\mathfrak{F}$, if %there exists a state $x_{\mathrm o}$ of $\Sigma$ such that 
$\Sigma$ is a minimal realization of $\mathfrak{F}$ from the zero initial state.
It can be shown that a LPV-SSA is a minimal realization of an input-output map, if and only if it is span-reachable and observable, moreover, all minimal LPV-SSA realizations of the same input-output map are isomorphic  \cite{petreczky2016}.
Furthermore, span-reachability and observability can be characterized by rank conditions of suitably defined matrices
\cite{petreczky2016}.

In this paper, in order to avoid excessive notation, we will make the following simplifying assumptions on the LPV-SSA models considered. 
\begin{Assumption}
\label{assum1}
In the sequel we assume that there is only one control input, i.e., $\NU=1$ and we consider only LPV-SSA models
of the form \eqref{equ:alpvss} for which the $D$ matrix is zero, and $C$ and $B$ matrices do not depend on the scheduling
parameters, i.e., $C(p)=C_0, B(p)=B_0$, $D(p)=0$ and hence $D_0=0$, $C_i=0,B_i=0, D_i=0$ for all $i=1,\ldots,\QNUM$,
\end{Assumption}

\section{Generalized transfer functions for LPV-SSA}
\label{sect:gentransfer}
Note that an input-output map $\mathfrak{F}$ of the form \eqref{equ:iofunction} is realizable by an LPV-SSA from the zero initial state satisfying Assumption \ref{assum1}, only if $\mathfrak{F}$ admits a so called
impulse response representation \cite{petreczky2016}, i.e., only if
 \begin{equation}
 \label{IIRsumC}
   \mathfrak{F}(u,p)(t)=\int_0^{t} (h_{\mathfrak{F}} \diamond p)(\delta,t) u(\delta)\ d\delta,
   \end{equation}
where for every $p \in \mathcal{P}$ the function 
$(h_{\mathfrak{F}} \diamond p)$ satisfies a number of technical conditions.
These conditions imply that $(h_{\mathfrak{F}} \diamond p)$
is an input-output map
induced by generating series in the sense of
\cite{Isi:Nonlin}, where the scheduling signal $p$
plays the role of the input. Recall from \cite[Chapter 3, Section 3.2]{Isi:Nonlin} that input-output maps which are induced by generating series also admit a Volterra-series representation. Moreover, if
$\mathfrak{F}$ has a realization by a LPV-SSA, then 
the input-output map w$p \mapsto (h_{\mathfrak{F}} \diamond p)$ can be realized by a bilinear system
whose matrices are matrices of the LPV-SSA realization of $\mathfrak{F}$. More precisely, by \cite{petreczky2016}
there exists a generating (Fliess) series 
$\theta_{\mathfrak{F}}: \mathfrak{S}(\mathbb{I}_0^{\QNUM}) \rightarrow \mathbb{R}^{\NY}$ defined on the set of all sequence 
of elements of $\mathbb{I}_0^{\QNUM}=\{0,1,\ldots,\QNUM\}$, such that 
\begin{align}
\label{lem:realiofunction:eq0.1}
\begin{split}
 & (h \diamond p)(\delta,t)=F_{\theta_{\mathfrak{F}}}[\sigma_{\delta} p](t-\delta)= \\ 
 & \theta_{\mathfrak{F}}(\epsilon) + 
 \sum_{k=1}^{\infty} \sum_{i_1,\ldots,i_k=0}^{\QNUM}
 \theta_{\mathfrak{F}}(i_1\cdots i_k)  \times  \\
 &  \times \int_{\delta}^{t} \int_{\delta}^{\tau_k} \cdots \int_{\delta}^{\tau_{2}}  p_{i_k}(\tau_k ) \cdots p_{i_1}(\tau_1) d\tau_k \cdots d\tau_{1}.
 \end{split}
 \vspace{-3mm}
\end{align}
Here, $p_0=1$,  $F_{\theta_{\mathfrak{F}}}$ denotes the input-output map
induced by the generating series $\theta_{\mathfrak{F}}$,
$\sigma_{\delta} p: s \mapsto p(\delta+s)$, and 
$F_{\theta_{\mathfrak{F}}}[\tilde{p}]$ is the value of the 
input-output map $F_{\theta_{\mathfrak{F}}}$ for the input signal
$\tilde{p}$. Here we use the standard notation used for generating (Fliess) series, see \cite{Isi:Nonlin}.  The second equation in \eqref{lem:realiofunction:eq0.1} is just the definition of an input-output map induced by a generating series. 
If $\Sigma$ is of the form \eqref{equ:alpvss}, satisfying Assumption \ref{assum1}, with $B(p)=B_0=B$ and $C(p)=C_0=C$, and 
$\Sigma$
is a realization of $\mathfrak{F}$, then it holds that
 \begin{align}\label{lem:realiofunction:eq0SISP}
 \vspace{-3mm}
 \theta_{\mathfrak{F} }(s) = C A_s B,
 \end{align}
where
for $s\!\!=\!\!\epsilon$, $A_s$ is the identity matrix, and for 
$s\!=\!s_1s_2 \cdots s_n $, $s_1,\ldots, s_n \!  \in \! \{0,\ldots n_{\mathrm p}\}$, $n \!>\! 0$, then $A_{s}=A_{s_n}A_{s_{n-1}} \cdots A_{s_1}$.

In other words, , the input-output map
$F_{\theta_{\mathfrak{F}}}$ is the input-output map of the bilinear system
\begin{equation}
\label{lpv2bilinSISO}
\vspace{-3mm}
\begin{split}
 & {\dot z}(t) = (A_0+\sum_{i=1}^{\QNUM} A_ip_i(t))z(t), ~ z(0)=B, \\
 & y(t)= C z(t),
\end{split} 
\vspace{-2mm}
\end{equation}
and hence, $y(t)=h_{\mathfrak{F}} \diamond p(\delta,t)$ is the output of the following
bilinear system at time $t$ 
\begin{equation}
\vspace{-2mm}
\label{lpv2bilin2SISO}
\begin{split}
 & {\dot z}(t) = (A_0+\sum_{i=1}^{\QNUM} A_i p_i(t))z(t), ~ z(\delta)=B, \\
 & y(t)= C z(t),
\end{split} 
\end{equation}
driven by the scheduling signal interpreted as input. 

Recall from \cite[Chapter 3, Section 3.2]{Isi:Nonlin} that 
input-output maps induced by generating series can also be represented by Volterra-kernels. For the specific case of $h_{\mathfrak{F}} \diamond p$,
this representation is as presented here; let us define
functions $W^{\mathfrak{F}}_{q_1,\ldots,q_k}(\tau_k,\tau_k,\ldots,\tau_0)$,
$\tau_{k} \ge \tau_{k-1} \ge \cdots \ge \tau_0 \ge 0$,
$q_1,\ldots,q_k \in \{1,\ldots,\QNUM\}$ and 
$W^{\mathfrak{F}}_{0}(\tau_0)$ as follows
\begin{equation}
\label{io:func:kernelSISO}
\begin{split}
 & W^{\mathfrak{F}}_{q_1 \cdots q_k}
 (\tau_{k},\tau_k,\ldots,\tau_0)= \\
 & \sum_{n_0,\ldots,n_k \in \mathbb{N}}
  \theta_{\mathfrak{F}}(0^{n_0}q_10^{n_1} \cdots q_k0^{n_k})
   \frac{\tau_0^{n_0}}{n_0 !} \Pi_{i=1}^{k} \frac{(\tau_i-\tau_{i-1})^{n_i}}{n_i !} \\
 &  W^{\mathfrak{F}}_{0}(\tau_0)=
 \sum_{n_0 \in \mathbb{N}}
  \theta_{i,j,\mathfrak{F},l}(0^{n_0}) \frac{\tau^{n_0}}{n_0 !}.
\end{split}   
\end{equation}
Here $0^k$ represents the $k$-fold repetition of the symbol $0$,
i.e., $0^0=\epsilon$, $0^{k}=\underbrace{00\cdots 0}_{k \mbox{times}}$.
It then follows that
\begin{equation*}
%\label{io:func:kernel4SISO}
\begin{split}
& (h_{\mathfrak{F}} \diamond p)(\delta,t) =
   W^{\mathfrak{F}}_{0}(t-\delta)+ \\
    & \sum_{k=1}^{\infty} \sum_{q_1\ldots q_k =1}^{\QNUM}
    \int_\delta^{t} \int_\delta^{\tau_k} \cdots \int_\delta^{\tau_1}
    W^{\mathfrak{F}}_{q_1,\ldots,q_k}(t-\delta,\tau_k-\delta,\ldots,\tau_1-\delta) \times  \\
   & p_{q_k}(\tau_k) \cdots p_{q_1}(\tau_1) d\tau_k\cdots d\tau_1 =  \\
   &  W^{\mathfrak{F}}_{0}(t-\delta)+ \\
    & \sum_{k=1}^{\infty} \sum_{q_1\ldots q_k =1}^{\QNUM}
    \int_0^{t-\delta} \int_0^{\tau_k} \cdots \int_0^{\tau_1}
    W^{\mathfrak{F}}_{q_1,\ldots,q_k}(t-\delta,\tau_k,\ldots,\tau_1) \times  \\
   & p_{q_k}(\tau_k+\delta) \cdots p_{q_1}(\tau_1+\delta)  d\tau_k\cdots d\tau_1.
\end{split}    
\end{equation*}
In particular, if $\Sigma$ is a realization of $\mathfrak{F}$
of the form \eqref{equ:alpvss}, with $C(p)=C$ and $B(p)=B$ being constants,
then 
\begin{equation*}
%\label{io:func:kernel5SISO}
\begin{split}
&  W^{\mathfrak{F}}_{0}(t)=Ce^{A_0t}B, \\
& W^{\mathfrak{F}}_{q_1,\ldots,q_k}(\tau_{k},\tau_{k-1},\ldots,\tau_0)= \\
 & Ce^{A_0(\tau_k-\tau_{k-1})}A_{q_k}e^{A_0(\tau_{k-1}-\tau_{k-2})}A_{q_{k-1}} \cdots e^{A_0(\tau_1-\tau_0)} A_{q_1}e^{A_{0}\tau_0}B
\end{split}    
\end{equation*}
The Volterra-kernels \eqref{io:func:kernelSISO} are the classical Volterra-kernels of input-affine nonlinear systems. In particular, 
we can take their multivariate  Laplace transforms 
resulting in a sequence of transfer functions 
$H_0(s)^{\mathfrak{F}}$, $H_{q_1,\ldots,q_k}^{\mathfrak{F}}(s_0,s_1,\ldots,s_k)$
\begin{align}
\label{io:func:kernelLaplaceSISO}
 & H^{\mathfrak{F}}_{q_1 \cdots q_k} (s_0,s_{1},s_{2} \ldots,s_k)=  \nonumber \\
 & \int_0^{\infty} \cdots \int_0^{\infty}
    W^{\mathfrak{F}}(\sum_{j=0}^{k} \tau_j, \sum_{j=0}^{k-1} \tau_j, \ldots,,\tau_0)e^{-(\sum_{j=1}^{k} s_j\tau_j)} d\tau_0\cdots d\tau_k, \nonumber \\
 &  H^{\mathfrak{F}}_{0}(s)=\int_0^{\infty} W^{\mathfrak{F}}_0(\tau)e^{-s\tau}d\tau.
\end{align}
Strictly speaking, the right-hand sides of in the equations \eqref{io:func:kernelLaplaceSISO} are well-defined only if
$\Re(s_i) > \sigma$ for a suitably chosen real number $\sigma$ which depends on $k$ and $q_1,\ldots,q_k$. In particular, if $A_0$ is stable, then $\sigma$ above can be taken to be $0$.
For the sake of simplicity, in the sequel we will implicitly assume that the functions $H_0^{\mathfrak{F}}$ and $H^{\mathfrak{F}}_{q_1,\ldots,q_k}$
are evaluated only for arguments for which the right-hand side of \eqref{io:func:kernelLaplaceSISO} is convergent.

In $\Sigma$ is a LPV-SSA realization of $\mathfrak{F}$
of the form \eqref{equ:alpvss} with $C(p)=C$ and $B(p)=B$, then
\begin{equation}
\label{io:func:kernelStateSpaceSISO}
\begin{split}
 & H^{\mathfrak{F}}_{q_1 \cdots q_k} (s_0,s_{1},s_{2} \ldots,s_k)= \\
   & C \Phi(s_k) A_{q_k} \Phi(s_{k-1}) A_{q_{k-1}}  \cdots A_{q_2}  \Phi(s_1) A_{q_1} \Phi(s_0) B, \\
 &  H^{\mathfrak{F}}_{0}(s_0)=C \Phi(s_0)  B,
\end{split}   
\end{equation}
where $\Phi(s) = (sI-A_0)^{-1}$ for all $s \in \IC$. 
\begin{Definition}[Generalized transfer functions]
The following sequence of transfer functions given by
\begin{equation}
\{H_0^{\mathfrak{F}},H_{q_1 \cdots q_k}^{\mathfrak{F}} \mid q_1,\ldots,q_k \in \{1,\ldots,n_{\mathrm p}\},k > 0\},
\end{equation}
is called the \emph{sequence of generalized transfer functions of $\mathfrak{F}$}. 
\end{Definition}

\section{The Loewner framework for modeling classical LTI systems}
\label{sect:lti}

In this section we present a brief overview of the Loewner framework, originally introduced in \cite{MA07}, for the LTI systems with multiple inputs and multiple outputs. For more details on various aspects of the method, we refer the reader to \cite{ALI17}. This framework is a data-driven modeling approach that constructs an LTI dynamical model with transfer function $\Htran_M: \IC \rightarrow \IC^{p \times m}$  which interpolates the given $2M$ samples (data measurements), for $M \in \mathbb{N}_+$. %Here, the data are assumed to be generated by a model with transfer function $\Htran: \IC \rightarrow \IC^{p \times m}$.
Let the left (or row) data values be given together with the right (or column) data values, as follows
\begin{equation}
\left.
\begin{array}{c}
(\mu_j,\bl_j^T,\bv_j^T) \\
\text{for $j=1,\dots,M$}
\end{array}
\right\}
\text{~~and~~}
\left\{
\begin{array}{c}
(\lambda_i,\br_i,\bw_i) \\
\text{for $i=1,\dots ,M$}
\end{array}
\right. ,
\label{eq:loewnerInput}
\end{equation}
where $\bv_j^T=\bl_j^T\Htran(\mu_j)$ and $\bw_i=\Htran(\lambda_i)\br_i$, with $\bl_j\in\IC^{p \times 1}$, $\br_i\in\IC^{m \times 1}$, $\bv_j\in\IC^{m \times 1}$ and $\bw_i\in\IC^{p \times 1}$. Then, split the distinct interpolation points $\{\eta_k\}_{k=1}^{2M} \subset \IC$ is split up into two disjoint subsets of same size, \ie
\begin{equation}\label{eq:shift}
\{\eta_k\}_{k=1}^{2M} =\{\mu_j\}_{j=1}^{M} \cup \{\lambda_i\}_{i=1}^{M}.
\end{equation}
The first step is to compute two matrices, i.e., the \emph{Loewner} matrix $\IL \in \IC^{M\times M}$ and \emph{shifted Loewner} matrix $\sIL \in \IC^{M\times M}$ defined for $i=1,\dots,M$ and $j=1,\dots,M$, as:
\begin{align}\label{eq:loewnerMatrices}
\begin{split}
[\IL]_{j,i} &= \dfrac{\bv_j^T\br_i - \bl_j^T\bw_i}{\mu_j - \lambda_i} 
= \dfrac{\bl_j^T\big( \Htran(\mu_j) - \Htran(\lambda_i) \big) \br_i}{\mu_j - \lambda_i}, \\
\,[\sIL]_{j,i} &= \dfrac{\mu_j\bv_j^T\br_i - \lambda_i\bl_j^T\bw_i}{\mu_j - \lambda_i}
= \dfrac{ \bl_j^T\big( \mu_j\Htran(\mu_j) - \lambda_i\Htran(\lambda_i) \big) \br_i}{\mu_j - \lambda_i}.
\end{split}
\end{align}
Additionally, we introduce the following matrices
\begin{align}
\IV = \left[ \begin{matrix}
\bv_1 &  \cdots & \bv_M
\end{matrix} \right]^T, \ \
\IW = \left[ \begin{matrix}
\bw_1  & \cdots & \bw_M
\end{matrix} \right],
\end{align}
with the following notation that holds for all $j,i=1,\ldots,M$ 
\begin{equation}
\bv_j^T = \bl^T_j H(\mu_j), \quad\mbox{and}\quad
\bw_i=H(\lambda_i)\br_i.
\end{equation}
%Finally, let
%\begin{align}
%\bM &= \text{diag}\left(\mu_1,\cdots,\mu_M \right), \ \ 
%\bLambda = \text{diag}\left(\lambda_1,\cdots,\lambda_M \right), \\ \bL &= \left[ \begin{matrix}
%\bl_1  & \cdots & \bl_M
%\end{matrix} \right], \ \  \bR = \left[ \begin{matrix}
%\br_1  & \cdots & \br_M
%\end{matrix} \right].
%\end{align}
Then, the Loewner LTI model $\Si_M$ is characterized by the following realization,
\begin{equation}\label{eq:loewnerDescrR}
\Si_M:
\begin{cases}
\bE_M   \dot{\bx}(t) = \bA_M \bx(t) + \bB_M \bu(t),\\
\hspace{5mm} \by(t) = \bC_M \bx(t),
\end{cases}
\end{equation}
where $\bE_M = -\IL$, $\bA_M = -\sIL$, $\bB_M = \IV$ and $\bC_M = \IW$. The transfer function of $\Si_M$ is given by
\begin{equation}
\Htran_M(s) = \bC_M(s \bE_M-\bA_M)^{-1} \bB_M.
\label{eq:loewnerDescrC}
\end{equation}

\begin{theorem}
Given the framework previously introduced, the function $\Htran_M$ interpolates $\Htran$ at the given driving frequencies and directions, i.e., for all $1\leq i \leq M$, it holds that
\begin{align}\label{eq:loewnerIntep}
\begin{split}
\bl_j^H\Htran_M(\mu_j) &= \bl_j^H\Htran(\mu_j), \\% \text{ and }
\Htran_M(\lambda_i)\br_i &= \Htran(\lambda_i) \br_i.
\end{split}
\end{align}
\end{theorem}

\vspace{2mm}

Next, we assume that the number of available measurements is larger than the underlying system's order denoted with $n$, i.e., $2M \geq n$. In this case, it was shown in \cite{MA07} that a minimal model $\Htran_n$ of dimension $n < M$ (that still interpolates the data) can be computed by means of projecting \eqref{eq:loewnerDescrR}. In order for this to be possible, the conditions below
\begin{equation}
%\scriptsize
\rank (\eta_k \IL - \sIL) = 
\rank ([\IL,\sIL]) = 
\rank ([\IL^T,\sIL^T]^T) = n,
\label{eq:rankCond}
\end{equation}
need to hold for $k=1,\ldots,2M$, where $\eta_k$ are as in \eqref{eq:shift}. In that case, let $\bY \in \IC^{M \times n}$ be the matrix containing the first $n$ left singular vectors of $[\IL,\sIL]$ and $\bX \in \IC^{M \times n}$  the matrix containing the first $n$ right singular vectors of $[\IL^T,\sIL^T]^T$. Then, construct a realization by means of projection as
\begin{align}    \label{eq:proj}
\begin{split}
\bE_n &= \bY^T \bE_M \bX,\,
\bA_n = \bY^T \bA_M \bX,\\
\bB_n &= \bY^T \bB_M,\,
\bC_n = \bC_M \bX,
\end{split}
\end{align}
which is equivalent to that in (\ref{eq:loewnerDescrR}).
%$\Htran_n$, given as, 
%\begin{equation}
%\Htran_n(s) = \bC_n( s \bE_n- \bA_n)^{-1} \bB_n,
%\label{eq:loewnerDescrCn}
%\end{equation}
The realization in (\ref{eq:proj}) encodes a \emph{minimal McMillan degree} equal to $\rank(\IL)$.

%The quadruple $\Hreal_n:(\bE_n,\bA_n,\bB_n,\bC_n,\bfz)$ is a descriptor realization of $\Htran_n$. Note that if $n$ in \eqref{eq:rankCond} is greater than $\rank(\IL)$, then $\Htran_n$ can either have a direct-feedthrough term or a polynomial part.

Finally, the number of singular vectors ($n$) that enter matrices $\bY$ and $\bX$ in (\ref{eq:proj}) could be indeed decreased to a value $r<n$. This would result in computing a reduced $r$-th order rational model that approximately interpolates the data. % denoted $\Htranr(\xi)=\hat C(\xi\hat E-\hat A)^{-1}\hat B$. 
This allows a trade-off between complexity of the resulting model and accuracy of  interpolation (as explained in \cite{MA07}).

\section{The proposed procedure}
\label{sect:main}
In what follows we describe the proposed procedure to construct a 
reduced order LPV-SSA $\tilde{\Sigma}=(\mathbb{P},\left\{ \tilde{A}_i, \tilde{B}_i, \tilde{C}_i, \tilde{D}_i \right\}_{i=0}^{\QNUM})$ from
an LPV-SSA of the form \eqref{equ:alpvss} satisfying Assumption \ref{assum1}.

To this end, let $N \in \mathbb{N}$ be a positive integer and introduce the following sequences of scalars:
\begin{enumerate}\label{disc_pts_param}
	\item $(\mu_0,\mu_1, \ldots,\mu_N)$ is the tuple of \textit{left interpolation points} in the frequency domain, with $\mu_i \in \IC$.
		\item $(\la_0,\la_1, \ldots,\la_N)$ is the tuple of \textit{right interpolation points} in the frequency domain with $\la_j \in \IC$.
		\item $(\ql_1 \ql_2 \ldots \ql_N)$ is the word of \textit{left expansion points} in the parameter domain with $\ql_i \in \mathbb{N}$.
		\item $(\qr_1\qr_2 \ldots\qr_N)$ is the word of \textit{right expansion points} in the parameter domain with $\qr_j \in \mathbb{N}$.
\end{enumerate}

 The associated {\textit generalized observability} matrix $\cO \in \IC^{(N+1) \times n_x}$ of the LPV-SSA \eqref{equ:alpvss} is put together as follows 
\begin{align}
\begin{split}
\cO&=\left[ \begin{array}{c} C \Phi(\mu_0) \\  C \Phi(\mu_0) A_{\ql_1} \Phi(\mu_1) \\  C \Phi(\mu_0) A_{\ql_1} \Phi(\mu_1) A_{\ql_2}  \Phi(\mu_2) \\
\vdots \\
C \Phi(\mu_0) A_{\ql_1} \Phi(\mu_1) A_{\ql_2}  \Phi(\mu_2) \cdots A_{\ql_N}  \Phi(\mu_N)
 \end{array} \right].
\end{split}
\end{align}
Recall that $C=C_0$, as by Assumption \ref{assum1} $C$ does not depend on the scheduling variable. 
Additionally,  the associated {\textit generalized controllability} matrix $\cR \in \IC^{n_x \times (N+1)}$ of \eqref{equ:alpvss} is also put together; below, we explicitly provide the entries of the $j$th column of matrix $\cR$, denote with $\cR_j$, as
\begin{align}
\mathcal{R}_1 &= \left[ \begin{array}{c} \Phi(\lambda_0)B
\end{array} \right], \ \ \mathcal{R}_2= \left[ \begin{array}{c}     \Phi(\lambda_1) A_{\qr_1} \Phi(\lambda_0)B 
\end{array} \right] \nonumber \\
\mathcal{R}_3 &= \left[ \begin{array}{c}\Phi(\lambda_2) A_{\qr_{2}} \Phi(\lambda_1)A_{\qr_1} \Phi(\lambda_0)B 
\end{array} \right] \\
\mathcal{R}_N &= \left[ \begin{array}{c}\Phi(\lambda_N) A_{\qr_{N}} \Phi(\lambda_{N-1})A_{\qr_{N-1}} \Phi(\lambda_{N-2})\cdots A_{\qr_1} \Phi(\lambda_0) B 
\end{array} \right]. \nonumber
\end{align}
Recall that $B=N_0$, as by Assumption \ref{assum1} $C$ does not depend on the scheduling variable. 
Then, put together a reduced-order model $\tilde{\Sigma}$ for the system in (\ref{equ:alpvss}), which is constructed from the original quantities. 
Define matrices for $1 \leq i \leq \QNUM$
\begin{align}\label{def_surr_model}
 \hat{E} = \cO \cR, \ \  \hat{A}_0 = \cO A_0 \cR, \ \  \hat{A}_i = \cO A_i \cR,  \  \hat{B} = \cO B, \  \hat{C} = C \cR.
\end{align}

%Note that (\ref{def_surr_model}) contains an additional matrix denoted with $\hat{E}$ - the model is in "descriptor format". For a realization in standard representation, 
Provided that $\hat{E}$ is nonsingular, one can write for all $1 \leq i \leq \QNUM$:
\begin{align}\label{def_surr_model2}
 \tilde{E} = I, \ \tilde{A}_0 = \hat{E}^{-1} \hat{A}_0, \  \tilde{A}_i = \hat{E}^{-1} \hat{A}_i,  \  \tilde{B} =  \hat{E}^{-1} \hat{B}, \  \tilde{C} = \hat{C}.
\end{align}

We then define the reduced order model $\tilde{\Sigma}$ as 
\begin{equation}
\label{def_surr_model3}
\tilde{\Sigma}=(\mathbb{P},\left\{ \tilde{A}_i, \tilde{B}_i, \tilde{C}_i, \tilde{D}_i \right\}_{i=0}^{\QNUM}),
\end{equation}
where $\tilde{D}_i=0$, $i=0,\ldots, n_{\mathrm p}$ and
$\tilde{C}_i=0$, $\tilde{B}_i=0$ for $i=1,\ldots, n_{\mathrm p}$
and $\tilde{B}_0=\tilde{B}$,$\tilde{C}=\tilde{C}_0$ and
$\{\tilde{A}_i\}_{i=1}^{n_{\mathrm p}}$, $\tilde{C},\tilde{B}$
are as in \eqref{def_surr_model2}.
%W_{k}=\bigcap_{q_1.. q_k} Kec C\Phi(\mu_0).A_{q_1}...... A_{q_k}\Phi(\mu_k)

%W_{k+1}=\phi(\mu_{k+1})^{-1} \bigcap_{q \in Q} A_q^{-1}W_k

%O_k=\bigcap_{q_1... q_k } Ker O_{q_1...q_k}(\mu_1....\mu_k)=W_1 \cap W_2 \cap ... \cap W_k

%u(t)=exp(s_0t), t=-\infty

%p_i(t)=exp(s_it)

%y(t)=\sum_k \sum_{q_1...q_k} H_{q_1...q_k}(s_k+...s_0,s_{k-1 }+..+s_0, ..,s_1+s_0,s_0)e^{s_0+...+s_kt}

%\mu_1... \mu_N fixed

\subsection{Data-driven interpretation}
\label{sec:dat_driv_int}

We will show in this section that the matrices computed in (\ref{def_surr_model}) can indeed be expressed in terms of samples of the transfer functions introduced in (\ref{io:func:kernelLaplaceSISO}).

For example, one can directly write the entries of vectors $\hat{B}= \cO B$ and $\hat{C}=  C \cR$ in (\ref{def_surr_model}) as
\begin{align}\label{hat_B_C}
  \hat{B} = \left[ \begin{matrix}
  H^{\mathfrak{F}}_{0} (\mu_0) \\
 H^{\mathfrak{F}}_{\ql_1} (\mu_{1},\mu_{0}) \\
  H^{\mathfrak{F}}_{\ql_2,\ql_1} (\mu_{2},\mu_1,\mu_{0}) \\
 \vdots 
  \end{matrix}
  \right], \   \hat{C} = \left[ \begin{matrix}
  H^{\mathfrak{F}}_{0} (\la_0) \\
 H^{\mathfrak{F}}_{\qr_1} (\la_{0},\la_{1}) \\
  H^{\mathfrak{F}}_{\qr_1,\qr_2} (\la_{0},\la_{1},\la_2) \\
 \vdots 
  \end{matrix}
  \right]^T
\end{align}
Additionally, the matrices $\bA_i$ for $1 \leq i \leq \QNUM$ are written element-wise, as follows:
\begin{align}\label{entries_hat_Ai}
    \left( \hat{A}_i \right)_{k+1,\ell+1} &= \cO_{k+1} A_i \cR_{\ell+1}  \\ &=  H^{\mathfrak{F}}_{\qr_1,\ldots,\qr_{\ell},i,\ql_{k},\ldots,\ql_{1}} (\la_{0},\ldots,\la_{\ell},\mu_{k},\ldots,\mu_{0}) \nonumber
\end{align}
Next, proceed to explicitly writing the $(k+1,\ell+1)$ entry of matrix $\hat{E}$ for $k,\ell \geq 0$. We make use of the recursion formulas on the rows  and columns of matrices $\cO$, and respectively, $\cR$ as (to have consistent notations, we enforce $\cO_0 = \cR_0 = 1$):
\begin{equation}
    \cO_{k+1} =  \cO_{k} A_{\ql_k} \Phi(\mu_k) , \ \   \cR_{\ell+1} = \Phi(\la_\ell) A_{\ql_\ell} \cR_{\ell}.
\end{equation}
Hence, based on the two identities presented above, we write the $(k+1,\ell+1)$ entry of matrix $\hat{E}$, for all $0 \leq k, \ell \leq N$, in the following way:
\begin{equation}
    \left( \hat{E} \right)_{k+1,\ell+1} = \cO_{k+1} I \cR_{\ell+1} = \cO_{k} A_{\ql_k} \Phi(\mu_k)  I \Phi(\la_\ell) A_{\qr_\ell} \cR_{\ell}.
\end{equation}
Next, we make use of the identity: $I = \frac{\Phi^{-1}(\mu_k)-\Phi^{-1}(\la_\ell)}{\mu_k-\la_\ell}$ and by substituting it in the equality above, it follows that:
\begin{align}\label{hat_E_kl}
    \left( \hat{E} \right)_{k+1,\ell+1} &= \cO_{k+1} I \cR_{\ell+1} \nonumber \\
    &= \cO_{k} A_{\ql_k} \Phi(\mu_k)  \frac{\Phi^{-1}(\mu_k)-\Phi^{-1}(\la_\ell)}{\mu_k-\la_\ell} \Phi(\la_\ell) A_{\qr_\ell} \cR_{\ell} \nonumber \\
    &= - \frac{\alpha_{k,\ell}- \beta_{k,\ell}}{\mu_k-\la_\ell},
\end{align}
where $\forall 0 \leq k, \ell \leq N$, \ and the following notations are used:
\begin{align}\label{alpha_beta_val}
\begin{split}
    \alpha_{k,\ell} &= \cO_{k} A_{\ql_k} \Phi(\mu_k) A_{\qr_\ell} \cR_{\ell} \\
    &=  H^{\mathfrak{F}}_{\qr_1,\ldots,\qr_{\ell},\ql_{k},\ldots,\ql_{1}} (\la_{0},\ldots,\la_{\ell-1},\mu_{k},\ldots,\mu_{0}),\\
    \beta_{k,\ell} &= \cO_{k} A_{\ql_k} \Phi(\la_\ell) A_{\qr_\ell} \cR_{\ell} \\
    &=  H^{\mathfrak{F}}_{\qr_1,\ldots,\qr_{\ell},\ql_{k},\ldots,\ql_{1}} (\la_{0},\ldots,\la_{\ell},\mu_{k-1},\ldots,\mu_{0}).
    \end{split}
\end{align}
Hence, we have shown that the entries of matrix $\hat{E}$ are divided differences composed of measurements corresponding to transfer functions in (\ref{io:func:kernelLaplaceSISO}). We proceed similarly for the entries of matrix $\hat{A}_0$. By using identity $A_0 = \frac{\mu_k\Phi^{-1}(\mu_k)-\la_\ell\Phi^{-1}(\la_\ell)}{\mu_k-\la_\ell}$, it follows
\begin{align}\label{hat_A0_kl}
    \left( \hat{A}_0 \right)_{k,\ell} &= \cO_k A_0 \cR_\ell = - \frac{\mu_k \alpha_{k,\ell}- \la_\ell \beta_{k,\ell}}{\mu_k-\la_\ell},
    \end{align}
where $\alpha_k$ and $\beta_\ell$ are as defined in (\ref{alpha_beta_val}), i.e., as samples of transfer functions introduced in (\ref{io:func:kernelLaplaceSISO}). So, we have shown that all matrices forming the data-driven surrogate realization in (\ref{def_surr_model}) are composed of transfer function measurements.

\begin{remark}
The matrix $\hat{E} \in \IC^{(N+1) \times (N+1)}$ defined element-wise as in (\ref{hat_E_kl}) is a Loewner matrix, while the matrix $\hat{A}_0 \in \IC^{(N+1) \times (N+1)}$ in (\ref{hat_A0_kl}) is a "shifted Loewner matrix", by following the terminology introduced in \cite{MA07}.
\end{remark}

\subsection{Interpolation property}

In this section we will show that the reduced model satisfies interpolation conditions. 

For the reduced-order LPV-SSA $\tilde{\Sigma}$  given in (\ref{def_surr_model3}), let
$\tilde{\mathfrak{F}}$ be the input-output map of
$\tilde{\Sigma}$. %introduce the following transfer functions
It then follows that
\begin{equation}
\label{io:func:kernelStateSpaceSISO_surr}
\begin{split}
 & H^{\tilde{\mathfrak{F}}}_{q_1 \cdots q_k} (s_{1},s_{2} \ldots,s_k)= \\
   & \tilde{C}  \widetilde{\Phi}(s_k) \tilde{A}_{q_k} \widetilde{\Phi}(s_{k-1}) A_{q_{k-1}}  \cdots \tilde{A}_{q_2}  \widetilde{\Phi}(s_1) A_{q_1} \widetilde{\Phi}(s_0) B, \\
 &  H^{\tilde{\mathfrak{F}}}_{0}(s_0)= \tilde{C} \widetilde{\Phi}(s_0)  \tilde{B},
\end{split}   
\end{equation}
where $\widetilde{\Phi}(s) = (sI-\tilde{A}_0)^{-1}$ for all $s \in \IC$. 

Given unit vectors $\bfe_{k+1}, \bfe_1 \in \IR^{N+1}$, one can write that:
\begin{align}
\begin{split}
  \bfe_{k+1}^T  \hat{E} \widetilde{\Phi}^{-1}(\la_0)  \bfe_1 &=   \bfe_{k+1}^T \hat{E} (\la_0I-\tilde{A}_0) \bfe_1 \\
  &=\bfe_{k+1}^T (\la_0 \hat{E}-\hat{A}_0) \bfe_1 \\
  &= \la_0 \bfe_{k+1}^T \hat{E} \bfe_1 - \bfe_{k+1}^T \hat{A}_0 \bfe_1  \\
  &= - \la_0  \frac{\alpha_{k,0}- \beta_{k,0}}{\mu_k-\la_0} + \frac{\mu_k \alpha_{k,0}- \la_0 \beta_{k,0}}{\mu_k-\la_0} \\
  &= \alpha_{k,0} =  \bfe_{k+1}^T \hat{B}=\bfe_{k+1}^T \hat{E} \tilde{B}
  \end{split}
\end{align}
Hence, we have shown that $ \bfe_{k+1}^T  \hat{E}\widetilde{\Phi}^{-1}(\la_0) \bfe_1 = \bfe_{k+1}^T \hat{E} \tilde {B}, \ \forall 0 \leq k \leq N$ which implies that
$\widetilde{\Phi}^{-1}(\la_0) \bfe_1 =  \tilde{B}$. By multiplying this identity to the left with $\tilde{C} \widetilde{\Phi}(\la_0)$, we can write that:
\begin{align}
\begin{split}
    &\tilde{C} \widetilde{\Phi}(\la_0) \widetilde{\Phi}^{-1}(\la_0) \bfe_1 =  \tilde{C} \widetilde{\Phi}(\la_0) \tilde{B} \Rightarrow  \tilde{C} \bfe_1 = \tilde{C} \widetilde{\Phi}(\la_0) \tilde{B} \\
    &\Rightarrow   H^{\tilde{\mathfrak{F}}}_{0} (\lambda_0) =   H^{\mathfrak{F}}_{0} (\lambda_0).
    \end{split}
\end{align}
Here we used that  $\tilde{C}=\hat{C} = C \cR$ and hence it follows that $\tilde{C}\bfe_1=C\cR\bfe_1=C\Phi(\lambda_0)B=H_0^{\mathfrak{F}}(\lambda_0)$, where $\Phi(s)=(sI-A_0)^{-1}$.
By repeating the above procedure, we can show that the interpolation condition $\hat{H}^{\mathfrak{F}}_{0} (\la_0) =   H^{\mathfrak{F}}_{0} (\la_0)$ also holds. In general, we can show that all measurements that appear as entries in the matrices of the reduced-order realization (\ref{def_surr_model3}), are actually matched by (\ref{io:func:kernelStateSpaceSISO_surr}). More precisely, we formulate the following result that explicitly states the interpolation conditions satisfied by the surrogate model.

\begin{theorem}
Given the framework previously introduced, the following $(N+1)^2+\QNUM(N+1)^2$ interpolation conditions are satisfied by the transfer functions in (\ref{io:func:kernelStateSpaceSISO_surr}):
\begin{equation}\label{interp_cond_Thrm2}
\begin{split}
       H^{\tilde{\mathfrak{F}}}_{\qr_1,\ldots,\qr_{\ell},\ql_{k},\ldots,\ql_{1}} (\la_{0},\ldots,\la_{\ell},\mu_{k-1},\ldots,\mu_{0}) \\ = H^{\mathfrak{F}}_{\qr_1,\ldots,\qr_{\ell},\ql_{k},\ldots,\ql_{1}} (\la_{0},\ldots,\la_{\ell},\mu_{k-1},\ldots,\mu_{0}), \\
      H^{\tilde{\mathfrak{F}}}_{\qr_1,\ldots,\qr_{\ell},\ql_{k},\ldots,\ql_{1}} (\la_{0},\ldots,\la_{\ell-1},\mu_{k},\ldots,\mu_{0}) \\ = H^{\mathfrak{F}}_{\qr_1,\ldots,\qr_{\ell},\ql_{k},\ldots,\ql_{1}} (\la_{0},\ldots,\la_{\ell-1},\mu_{k},\ldots,\mu_{0}), \\
    H^{\tilde{\mathfrak{F}}}_{\qr_1,\ldots,\qr_{\ell},i,\ql_{k},\ldots,\ql_{1}} (\la_{0},\ldots,\la_{\ell},\mu_{k},\ldots,\mu_{0}) \\ = H^{\mathfrak{F}}_{\qr_1,\ldots,\qr_{\ell},i,\ql_{k},\ldots,\ql_{1}} (\la_{0},\ldots,\la_{\ell},\mu_{k},\ldots,\mu_{0}),
      \end{split}
\end{equation}
 for all $0 \leq k, \ell \leq N$ and $1 \leq i \leq \QNUM$.
\end{theorem}

\begin{example}
Below, we illustrate the proposed extension of the Loewner framework through one simple example ($\QNUM = 1$ and $N=2$).
%\begin{equation}
%\begin{split}
%& H^{\mathfrak{F}}_{11} (s_{0},s_1,s_2)= C \Phi(s_2)  A_{1} \Phi(s_1) A_{1} \Phi(s_0) B, \\
% & H^{\mathfrak{F}}_{1} (s_{0},s_1)= C \Phi(s_1) A_{1} \Phi(s_0) B, \ \  H^{\mathfrak{F}}_{0}(s_0)=C \Phi(s_0) B.
%\end{split}   
%\end{equation}
The associated {\textit generalized observability and controllability} matrices are put together as follows 
\begin{align}
\begin{split}
		\cO&=\left[ \begin{array}{c} C \Phi(\mu_0) \\  C \Phi(\mu_0) A_1 \Phi(\mu_1)  \end{array} \right], \\
		\mathcal{R} &= \left[ \begin{array}{cc} \Phi(\lambda_0)B &    \Phi(\lambda_1) A_1 \Phi(\lambda_0)B \end{array} \right].
		\end{split}
\end{align}
The next step is to show that we can interpret matrices:
\begin{align*}
   \hat{E} = \cO \cR, \ \hat{A}_0 = \cO A_0 \cR, \ \hat{A}_1 =  \cO A_1 \cR, \ \hat{B} = \cO B, \ \hat{C} =  C \cR,
\end{align*}
in terms of data, i.e., measurements of transfer functions. To do so, we repeat the general procedure presented in Section \ref{sec:dat_driv_int} for this simplified scenario, and hence write that
\small
\begin{align}
\begin{split}
   \hat{E} &= -\left[ \begin{matrix} \frac{H^{\mathfrak{F}}_{0} (\mu_0)-H^{\mathfrak{F}}_{0} (\la_0)}{\mu_0-\la_0} & \frac{H^{\mathfrak{F}}_{1} (\la_0,\mu_0)-H^{\mathfrak{F}}_{1} (\la_0,
  \la_1)}{\mu_0-\la_1} \\[2mm]
  \frac{H^{\mathfrak{F}}_{1} (\mu_1,\mu_0)-H^{\mathfrak{F}}_{1} (\la_0,
  \mu_0)}{\mu_1-\la_0} & \frac{H^{\mathfrak{F}}_{1,1} (\la_0,\mu_1,\mu_0)-H^{\mathfrak{F}}_{1,1} (\la_0,\la_1,
  \mu_0)}{\mu_1-\la_1}
    \end{matrix}
    \right], \\
       \hat{A}_0 &= -\left[ \begin{matrix}  \frac{\mu_0 H^{\mathfrak{F}}_{0} (\mu_0)-\la_0H^{\mathfrak{F}}_{0} (\la_0)}{\mu_0-\la_0} & \frac{\mu_0 H^{\mathfrak{F}}_{1} (\la_0,\mu_0)-\la_1 H^{\mathfrak{F}}_{1} (\la_0,
  \la_1)}{\mu_0-\la_1} \\[2mm]
  \frac{\mu_1 H^{\mathfrak{F}}_{1} (\mu_1,\mu_0)-\la_0 H^{\mathfrak{F}}_{1} (\la_0,
  \mu_0)}{\mu_1-\la_0} & \frac{\mu_1 H^{\mathfrak{F}}_{1,1} (\la_0,\mu_1,\mu_0)-\la_1 H^{\mathfrak{F}}_{1,1} (\la_0,\la_1,
  \mu_0)}{\mu_1-\la_1}
    \end{matrix}
    \right], \\
    \hat{A}_1 &= \left[ \begin{matrix}
    H^{\mathfrak{F}}_{1} (\la_0,\mu_0) & H^{\mathfrak{F}}_{1,1} (\la_0,\la_1,\mu_0) \\
    H^{\mathfrak{F}}_{1,1} (\la_0,\mu_1,\mu_0) & H^{\mathfrak{F}}_{1,1,1} (\la_0,\la_1,\mu_1,\mu_0)
    \end{matrix}
    \right], \\
        \hat{B} &= \left[ \begin{matrix}
    H^{\mathfrak{F}}_{0} (\mu_0) \\ H^{\mathfrak{F}}_{1} (\mu_1,\mu_0) 
    \end{matrix} \right], \ \         \hat{C} = \left[ \begin{matrix}
    H^{\mathfrak{F}}_{0} (\la_0) & H^{\mathfrak{F}}_{1} (\la_0,\la_1) 
    \end{matrix} \right].
    \end{split}
   \end{align}
\normalsize
So, in this simple case in which $N=\QNUM=1$, it follows that $(N+1)^2+\QNUM(N+1)^2 = 8$ interpolation conditions are satisfied by the reduced model $\tilde{\Sigma}$ calculated according to \eqref{def_surr_model3} and \eqref{def_surr_model2}. Below, we enumerate the transfer function values that are matched:
\begin{align*}
     &H^{\mathfrak{F}}_{0} (\mu_0), \     H^{\mathfrak{F}}_{0} (\la_0), \     H^{\mathfrak{F}}_{1} (\mu_1,\mu_0), \     H^{\mathfrak{F}}_{1} (\la_0,\mu_0), \ \     H^{\mathfrak{F}}_{1} (\la_0,\la_1), \\ &H^{\mathfrak{F}}_{1,1} (\la_0,\mu_1,\mu_0), \     H^{\mathfrak{F}}_{1,1} (\la_0,\la_1,\mu_0), \ \ H^{\mathfrak{F}}_{1,1,1} (\la_0,\la_1,\mu_1,\mu_0).
\end{align*}
%Then, let's analyze the case $\QNUM = 2$. For $q_1,q_2 \in \{1,2\}$
%\begin{equation}
%\begin{split}
%& H^{\mathfrak{F}}_{q_1,q_2} (s_{0},s_1,s_2)= C \Phi(s_2) A_{q_2} \Phi(s_1) A_{q_1} \Phi(s_0) B, \\
% & H^{\mathfrak{F}}_{q_1} (s_{0},s_1)= C \Phi(s_1) A_{q_1} \Phi(s_0) B, \ \ H^{\mathfrak{F}}_{0}(s_0)=C \Phi(s_0) B.
%\end{split}   
%\end{equation}

%Let $\Phi(s) = (sI-A_0)^{-1}$ for all $s \in \IC$. The associated {\textit generalized observability and controllability} matrices are computed as follows 
%\begin{align}
%\begin{split}
%		\cO&=\left[ \begin{array}{c} C \Phi(\mu_0) \\  C \Phi(\mu_0) A_1 \Phi(\mu_1) \\  C \Phi(\mu_0) A_1 \Phi(\mu_1) A_2  \Phi(\mu_2) \end{array} \right], \\
%		\mathcal{R} &= \left[ \begin{array}{ccc} \Phi(\lambda_0)B &    \Phi(\lambda_1) A_1 \Phi(\lambda_0)B & \Phi(\lambda_2) A_2 \Phi(\lambda_1) A_1 \Phi(\lambda_0)B \end{array} \right].
%		\end{split}
%\end{align}
%The next step is to show that we can interpret matrices:
%\begin{align}
%    \cO \cR, \ \cO A_0 \cR, \ \cO A_1 \cR,  \ \cO A_2 \cR, \ \cO B, \  C \cR,
%\end{align}
%in terms of data, i.e., measurements of transfer functions.
\end{example}

\section{Numerical example}
\label{sect:num}

\begin{figure}[ht]
\hspace{-4mm}		
	\includegraphics[width=1.1\linewidth]{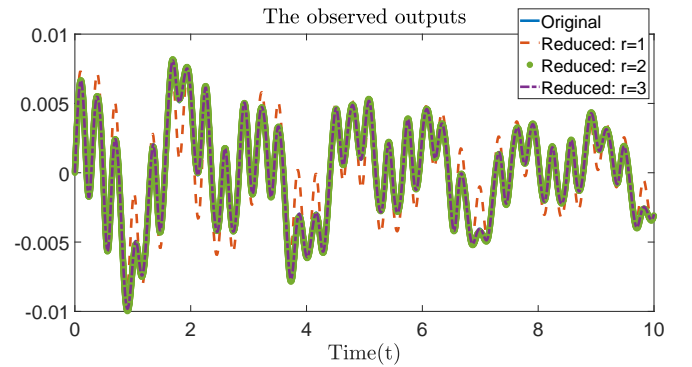}
	\vspace{-6mm}
	\caption{The observed outputs (original and reduced).}
	\label{fig:1}
	\vspace{-3mm}
\end{figure}

In this section we revisit the example presented in \cite{petreczky2016} (Section III, Example 1). Based on Assumption (\ref{assum1}) that was imposed in Section \ref{sec:pre} of the current paper, the B and C matrices will be considered to be constant. Additionally, we shift the original matrix $A_0$ from \cite{petreczky2016} so that all its poles are located into the left-half (complex) plane. Finally, choose $\QNUM = 2$ (originally, $\QNUM = 1$ was enforced) and assume zero initial conditions. The system matrices of the modified system are given as follows:
\begin{align}
    A_0 &= \left[ \begin{matrix}
    -1 & 1 & -1\\ -1 & -2 & 1\\ -1 & 1 & -3
    \end{matrix} \right], \     A_1 = \left[ \begin{matrix}
   1 & -1 & -1\\ -1 & 2 & 0\\ -1 & 0 & 2
    \end{matrix} \right], \ B_0 =  \left[ \begin{matrix}
  1\\ 0\\ 0
    \end{matrix} \right], \nonumber \\
    A_2 &= \left[ \begin{matrix}
   0 & -1 & 1\\ 0 & 1 & 2\\ 2 & 1 & 0
    \end{matrix} \right], \ \ C_0 = \left[ \begin{matrix}
  1 & -1 & -1
    \end{matrix} \right], \ \ D = 0.
    \vspace{-2mm}
\end{align}
The control input is chosen as $u(t) = 0.1 \cos(20t) \cdot e^{-0.1 t}$, while the scheduling signals are purely oscillatory, with different main frequencies, i.e., $p_1(t) = 2.5 \sin(5 \pi t)$ and $p_2(t) = 1.25 \sin(7 \pi t)$. We apply the newly-proposed method for $N=2$ and the following choice of left and right interpolation points (located on the imaginary axis; here $\imath = \sqrt{-1}$).
\begin{align}
  \begin{cases}  \mu_0 = 2 \imath, \  \mu_1 = 4 \imath, \  \mu_2 = 6 \imath, \\
  \la_0 = 3 \imath, \  \la_1 = 5 \imath, \  \la_3 = 8 \imath.
  \end{cases}
\end{align}
It is to be noted that we construct three reduced-order models of dimension $r$ for all values $1 \leq r \leq 3$, by following the procedure outlined in Section \ref{sect:main}. The accuracy of these interpolation-based surrogate models is tested by means of time-domain simulations. We simulate the original system together with the three reduced ones on a time range of $[0,10]$s (by applying a classical first-order Euler scheme on $5 \cdot 10^4$ points). The observed outputs of the original system, together with the outputs of the three reduced models are depicted in Fig.\;\ref{fig:1}.

\begin{figure}[ht]
\hspace{-2mm}		
	\includegraphics[width=1.1\linewidth]{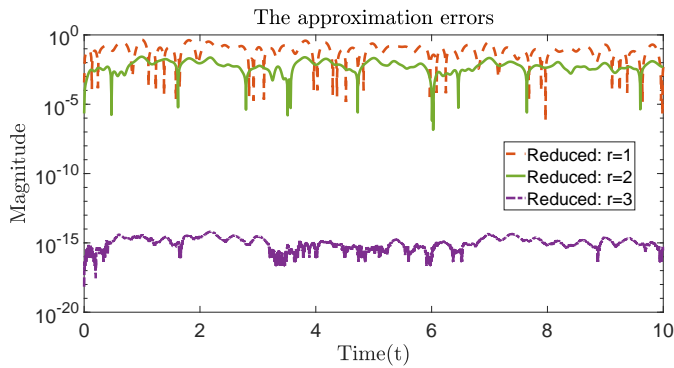}
	\vspace{-6mm}
	\caption{The relative approximation errors.}
	\label{fig:2}
	\vspace{-3mm}
\end{figure}

Additionally, we compute the magnitude of the relative approximation error for each reduced dimension $r \in \{1,2,3\}$ and depict the curves in Fig. \ref{fig:2}. Clearly, the order $r = 3$ system computed by means of the new method perfectly matches the response of the original system (the approximation errors are in the range of machine precision). The other two reduced systems, of course enforce higher errors; in particular, the output of the one of order $r = 2$ follows quite accurately the original response (as illustrated in Fig.\;\ref{fig:1}).

\section{Conclusion}
\label{sect:concl}
We have proposed an extension of the  Loewner framework to LPV systems with an affine dependence on parameters. The proposed framework yields a model reduction procedure which is based on matching the frequency response of the original system at some particular frequencies. In order to avoid complex notations, we have restricted the attention to the single input case and to models for which the $B$ and $C$ matrices do not depend on the scheduling parameters. Moreover, we analyzed a particular choice of frequencies to be matched. Future research will be directed towards extending these results to general LPV systems with affine dependence on parameters. Other research directions include finding system theoretic interpretations for the proposed method, i.e., showing that for certain inputs and scheduling signals the time-domain responses of the original and reduced model coincide, possibly after filtering. Finally, we plan to test the proposed method for more complex models. 

\small
\bibliographystyle{IEEEtran}
\bibliography{Bibliography,Loew_ref,petreczky_chapter}

\begin{thebibliography}{10}
\providecommand{\url}[1]{#1}
\csname url@samestyle\endcsname
\providecommand{\newblock}{\relax}
\providecommand{\bibinfo}[2]{#2}
\providecommand{\BIBentrySTDinterwordspacing}{\spaceskip=0pt\relax}
\providecommand{\BIBentryALTinterwordstretchfactor}{4}
\providecommand{\BIBentryALTinterwordspacing}{\spaceskip=\fontdimen2\font plus
\BIBentryALTinterwordstretchfactor\fontdimen3\font minus
  \fontdimen4\font\relax}
\providecommand{\BIBforeignlanguage}[2]{{%
\expandafter\ifx\csname l@#1\endcsname\relax
\typeout{** WARNING: IEEEtran.bst: No hyphenation pattern has been}%
\typeout{** loaded for the language `#1'. Using the pattern for}%
\typeout{** the default language instead.}%
\else
\language=\csname l@#1\endcsname
\fi
#2}}
\providecommand{\BIBdecl}{\relax}
\BIBdecl

\bibitem{Rugh00}
W.~Rugh and J.~S. Shamma, ``Research on gain scheduling,'' \emph{Automatica},
  vol.~36, no.~10, pp. 1401--1425, 2000.

\bibitem{LPVBook2012}
J.~Mohammadpour and C.~W. Scherer, \emph{Control of Linear Parameter Varying
  Systems with Applications}.\hskip 1em plus 0.5em minus 0.4em\relax
  Heidelberg: Springer, 2012.

\bibitem{Val13b}
D.~Vizer, G.~Merc{\`e}re, O.~Prot, E.~Laroche, and M.~Lovera, ``Linear
  fractional {LPV} model identification from local experiments: an
  {$H_{\infty}$}-based optimization technique,'' in \emph{In {IEEE} Conference
  on Decision and Control}, Florence, Italy, December 2013.

\bibitem{Toth2010SpringerBook}
R.~T{\'o}th, ``Modeling and identification of linear parameter-varying
  systems,'' in \emph{Lecture Notes in Control and Information Sciences, Vol.
  403}.\hskip 1em plus 0.5em minus 0.4em\relax Heidelberg: Springer, 2010.

\bibitem{toth12}
R.~T\'{o}th, H.~S. Abbas, and H.~Werner, ``On the state-space realization of
  {LPV} input-output models: Practical approaches,'' \emph{{IEEE} Trans. Contr.
  Syst. Technol.}, vol.~20, pp. 139--153, Jan. 2012.

\bibitem{Giarre2002}
B.~Bamieh and L.~Giarr{\'e}, ``Identification of linear parameter varying
  models,'' \emph{International Journal of Robust and Nonlinear Control},
  vol.~12, pp. 841--853, 2002.

\bibitem{Wingerden09}
J.~W. {van Wingerden} and M.~Verhaegen, ``Subspace identification of bilinear
  and {LPV} systems for open- and closed-loop data,'' \emph{Automatica},
  vol.~45, no.~2, pp. 372--381, 2009.

\bibitem{LopesDosSantos2008}
P.~L. {dos Santos}, J.~A. Ramos, and J.~L.~M. {de Carvalho}, ``Identification
  of {LPV} systems using successive approximations,'' in \emph{Proc. of 47th
  {IEEE} Conference on Decision and Control}, 2008, pp. 4509--4515.

\bibitem{Sznaier:01}
M.~Sznaier and C.~Mazzaro, ``An {LMI} approach to the identification and
  (in)validation of {LPV} systems,'' in \emph{Perspectives in robust control.
  Lecture Notes in Control and Information Sciences}, S.~Moheimani, Ed.\hskip
  1em plus 0.5em minus 0.4em\relax London: Springer, 2001, vol. 268, pp.
  327--346.

\bibitem{Verdult02}
V.~Verdult and M.~Verhaegen, ``Subspace identification of multivariable linear
  parameter-varying systems,'' \emph{Automatica}, vol.~38, no.~5, pp. 805--814,
  2002.

\bibitem{BlanchiniTAC}
F.~Blanchini, D.~Casagrande, S.~Miani, and U.~Viaro, ``Stable {LPV} realization
  of parametric transfer functions and its application to gain-scheduling
  control design,'' \emph{IEEE Transactions on Automatic Control}, vol.~55,
  no.~10, pp. 2271--2281, 2010.

\bibitem{Ant05}
A.~C. Antoulas, \emph{Approximation of large-scale dynamical systems}, ser.
  Advances in Design and Control.\hskip 1em plus 0.5em minus 0.4em\relax SIAM,
  2005.

\bibitem{BGW15surveyMOR}
P.~Benner, S.~Gugercin, and K.~Willcox, ``A survey of projection-based model
  reduction methods for parametric dynamical systems,'' \emph{SIAM Review},
  vol.~57, no.~4, pp. 483--531, 2015.

\bibitem{AntBG20}
A.~C. Antoulas, C.~Beattie, and S.~G\"{u}\u{g}ercin, \emph{Interpolatory
  methods for model reduction}, ser. Computational Science and Engineering
  21.\hskip 1em plus 0.5em minus 0.4em\relax SIAM, Philadelphia, 2020.

\bibitem{farhood2003}
M.~Farhood, C.~Beck, and G.~Dullerud, ``On the model reduction of nonstationary
  {LPV} systems,'' in \emph{Proc. of the American Control Conference (ACC)},
  Denver, CO, USA, Jun. 2003, pp. 3869 -- 3874.

\bibitem{dehillerin2011}
S.~D. Hillerin, G.~Scorletti, and V.~Fromion, ``Reduced-complexity controllers
  for {LPV} systems: Towards incremental synthesis,'' in \emph{Proc. of the
  50th IEEE Conference on Decision and Control and European Control Conference
  (CDC-ECC)}, Orlando, FL, USA, Dec. 2011, pp. 3404 -- 3409.

\bibitem{adegas2013}
F.~Adegas, I.~Sonderby, M.~Hansen, and J.~Stoustrup, ``Reduced-order lpv model
  of flexible wind turbines from high fidelity aeroelastic codes,'' in
  \emph{Proc. of the IEEE International Conference on Control Applications
  (CCA)}, Hyderabad, Aug. 2013, pp. 424 -- 429.

\bibitem{wood1996}
G.~Wood, P.~Goddard, and K.~Glover, ``Approximation of linear parameter-varying
  systems,'' in \emph{Proc. of the 35th IEEE Conference on Decision and
  Control}, Kobe, Dec. 1996, pp. 406 -- 411.

\bibitem{widowatin}
Widowati, R.~Bambang, R.~Saragih, and S.~Nababan, ``Model reduction for
  unstable {LPV} systems based on coprime factorizations and singular
  perturbation,'' in \emph{Proc. of the 5th Asian Control Conference},
  Melbourne, Jul. 2004, pp. 963 -- 970.

\bibitem{MertBastug:CDC2015}
M.~Bastug, M.~Petreczky, R.~T\'oth, R.~Wisniewski, J.~Leth, and D.~Efimov,
  ``Moment matching based model reduction for lpv state-space models,'' in
  \emph{Decision and Control (CDC), 2015 IEEE 54rd Annual Conference on}, 2015.

\bibitem{BennerLPV}
P.~Benner, X.~Cao, and W.~Schilders, ``A bilinear {H}$_2$ model order reduction
  approach to linear parameter-varying systems,'' \emph{Adv Comput Math},
  vol.~45, pp. 2241--2271, 2019.

\bibitem{TothModRed}
S.~Z. {Rizvi}, J.~{Mohammadpour}, R.~{Tóth}, and N.~{Meskin}, ``A kernel-based
  {PCA} approach to model reduction of linear parameter-varying systems,''
  \emph{IEEE Transactions on Control Systems Technology}, vol.~24, no.~5, pp.
  1883--1891, 2016.

\bibitem{SzaboModelRed}
T.~Luspay, T.~P\'eni, I.~G\"ozse, Z.~Szab\'o, and B.~Vanek, ``Model reduction
  for {LPV} systems based on approximate modal decomposition,''
  \emph{International Journal for Numerical Methods in Engineering}, vol. 113,
  no.~6, pp. 891--909, 2018.

\bibitem{WeilandModelRed}
S.~{Schouten}, D.~{Lou}, and S.~{Weiland}, ``Model reduction for linear
  parameter-varying systems through parameter projection,'' in \emph{2019 IEEE
  58th Conference on Decision and Control (CDC)}, 2019, pp. 7800--7805.

\bibitem{MA07}
A.~Mayo and A.~Antoulas, ``A framework for the solution of the generalized
  realization problem,'' \emph{Linear Algebra and Its Applications}, vol. 425,
  no. 2-3, pp. 634--662, 2007.

\bibitem{VictorLoewner}
I.~V. Gosea, M.~Petreczky, and A.~C. Antoulas, ``Data-driven model order
  reduction of linear switched systems in the {L}oewner framework,'' \emph{SIAM
  Journal on Scientific Computing}, vol.~40, no.~2.

\bibitem{AGI16}
A.~C. Antoulas, I.~V. Gosea, and A.~C. Ionita, ``Model reduction of bilinear
  systems in the {L}oewner framework,'' \emph{SIAM Journal on Scientific
  Computing}, vol. 38(5), pp. B889--B916, 2016.

\bibitem{Rugh96book}
W.~J. Rugh, \emph{Linear System theory}.\hskip 1em plus 0.5em minus 0.4em\relax
  Prentice-Hall, 1996.

\bibitem{SirajCDC2012}
M.~Siraj, R.~Toth, and S.~Weiland, ``Joint order and dependency reduction for
  {LPV} state-space models,'' in \emph{Decision and Control (CDC), 2012 IEEE
  51st Annual Conference on}, 2012, pp. 6291--6296.

\bibitem{petreczky2016}
M.~Petreczky, G.~Merc{\`{e}}re, and R.~T{\'{o}}th, ``{Affine LPV systems :
  realization theory , input-output equations and relationship with linear
  switched systems},'' \emph{IEEE Transactions on Automatic Control}, vol.~62,
  pp. 4667--4674, 2017.

\bibitem{WebsterBook}
R.~Webster, \emph{Convexity}.\hskip 1em plus 0.5em minus 0.4em\relax Oxford,
  1994.

\bibitem{Isi:Nonlin}
A.~Isidori, \emph{Nonlinear Control Systems}.\hskip 1em plus 0.5em minus
  0.4em\relax Springer Verlag, 1989.

\bibitem{ALI17}
A.~C. Antoulas, S.~Lefteriu, and A.~C. Ionita, ``A tutorial introduction to the
  {L}oewner framework for model reduction,'' in \emph{Model Reduction and
  Approximation}.\hskip 1em plus 0.5em minus 0.4em\relax SIAM, 2017, ch.~8, pp.
  335--376.

\end{thebibliography}


\begin{thebibliography}{10}
\providecommand{\url}[1]{#1}
\csname url@samestyle\endcsname
\providecommand{\newblock}{\relax}
\providecommand{\bibinfo}[2]{#2}
\providecommand{\BIBentrySTDinterwordspacing}{\spaceskip=0pt\relax}
\providecommand{\BIBentryALTinterwordstretchfactor}{4}
\providecommand{\BIBentryALTinterwordspacing}{\spaceskip=\fontdimen2\font plus
\BIBentryALTinterwordstretchfactor\fontdimen3\font minus
  \fontdimen4\font\relax}
\providecommand{\BIBforeignlanguage}[2]{{%
\expandafter\ifx\csname l@#1\endcsname\relax
\typeout{** WARNING: IEEEtran.bst: No hyphenation pattern has been}%
\typeout{** loaded for the language `#1'. Using the pattern for}%
\typeout{** the default language instead.}%
\else
\language=\csname l@#1\endcsname
\fi
#2}}
\providecommand{\BIBdecl}{\relax}
\BIBdecl

\bibitem{Val13b}
D.~Vizer, G.~Merc{\`e}re, O.~Prot, E.~Laroche, and M.~Lovera, ``Linear
  fractional {LPV} model identification from local experiments: an
  {$H_{\infty}$}-based optimization technique,'' in \emph{Proc. of 52nd
  IEEEConference on Decision and Control}, 2013, pp. 4559 -- 4564.

\bibitem{CoxSubspace}
P.~B. Cox, R.~T{\'o}th, and M.~Petreczky, ``Estimation of {LPV-SS} models with
  static dependency using correlation analysis,'' in \emph{1st IFAC Workshop on
  Linear Parameter Varying Systems}, 2015, pp. 91--96.

\bibitem{MertLPVCDC2015}
M.~Bastug, M.~Petreczky, R.~T\'oth, R.~Wisniewski, J.~Leth, and D.~Efimov,
  ``Moment matching based model reduction for {LPV} state-space models,'' in
  \emph{Proc. of 54th IEEE Conference on Decision and Control}, 2015, pp. 5334
  -- 5339.

\bibitem{BelikovSCL2014}
J.~Belikov, U.~Kotta, and M.~T\~onso, ``Comparison of {LPV} and nonlinear
  system theory: A realization problem,'' \emph{Systems \& Control Letters},
  vol.~64, pp. 72 -- 78, 2014.

\bibitem{MoogBook}
G.~Conte, C.~H. Moog, and A.~M. Perdon, \emph{Algebraic Methods for Nonlinear
  Control Systems.}\hskip 1em plus 0.5em minus 0.4em\relax Springer-Verlag,
  2007.

\bibitem{PM12}
M.~Petreczky and G.~Merc{\`e}re, ``Affine {LPV} systems: realization theory,
  input-output equations and relationship with linear switched systems,'' in
  \emph{Proc. of 51th {IEEE} Conference on Decision and Control}, 2012, pp.
  4511 -- 4516.

\bibitem{MPGM:LPVTech}
M.~Petreczky, G.~Merc\`ere, and R.~T\'oth, ``Affine {LPV} systems: realization
  theory, input-output equations and relationship with linear switched
  systems,'' Tech. Rep., 2012, arxive 1209.0345.

\bibitem{LPVKalmanPaperArxive}
M.~Petreczky, R.~T\'oth, and G.~Merc\`ere, ``Realization theory of {LPV}
  state-space representations with affine dependence,'' Tech. Rep., 2016, arxiv
  1601.02777v2.

\bibitem{Pet11}
M.~Petreczky and J.~{van Schuppen}, ``Partial realization theory for linear
  switched systems: a formal power series approach,'' \emph{Automatica},
  vol.~47, pp. 2177--2184, 2011.

\bibitem{Pet12}
M.~Petreczky, L.~Bako, and J.~{van Schuppen}, ``Realization theory of
  discrete-time linear switched system,'' \emph{Automatica}, vol.~49, no.~11,
  pp. 3337--3344, 2013.

\bibitem{PetCocv11}
M.~Petreczky, ``Realization theory for linear and bilinear switched systems:
  formal power series approach - part i: realization theory of linear switched
  systems,'' \emph{ESAIM Control, Optimization and Calculus of Variations},
  vol.~17, pp. 410--445, 2011.

\bibitem{WebsterBook}
R.~Webster, \emph{Convexity}.\hskip 1em plus 0.5em minus 0.4em\relax Oxford,
  1994.

\bibitem{Isi:Nonlin}
A.~Isidori, \emph{Nonlinear Control Systems}.\hskip 1em plus 0.5em minus
  0.4em\relax Springer Verlag, 1989.

\bibitem{ALI17}
A.~C. Antoulas, S.~Lefteriu, and A.~C. Ionita, ``A tutorial introduction to the
  {L}oewner framework for model reduction,'' in \emph{Model Reduction and
  Approximation}.\hskip 1em plus 0.5em minus 0.4em\relax SIAM, 2017, ch.~8, pp.
  335--376.

\bibitem{Antoulas16}
A.~Antoulas, ``The {L}oewner framework and transfer functions of
  singular/rectangular systems,'' \emph{App. Math. Letters}, vol.~54, pp.
  36--47, 2016.

\bibitem{MA07}
A.~Mayo and A.~Antoulas, ``A framework for the solution of the generalized
  realization problem,'' \emph{Linear Algebra and Its Applications}, vol. 425,
  no. 2-3, pp. 634--662, 2007.

\end{thebibliography}

\end{document}